\documentclass[reprint,onecolumn,notitlepage,prd,amsfonts,amssymb,noshowpacs,nofootinbib,preprintnumbers]{revtex4-1}
\usepackage{amsmath,graphicx,hyperref,epsfig,color}
\usepackage[normalem]{ulem}

\begin{document}

\title{The two faces of mimetic Horndeski gravity: disformal transformations and Lagrange multiplier}
\author{
Frederico Arroja$^{1}$\footnote{arroja@phys.ntu.edu.tw}, Nicola Bartolo$^{2,3}$\footnote{nicola.bartolo@pd.infn.it}, Purnendu Karmakar$^{2,3}$\footnote{purnendu.karmakar@pd.infn.it} and Sabino Matarrese$^{2,3,4}$\footnote{sabino.matarrese@pd.infn.it}
}
\affiliation{
{}$^{1}$ Leung Center for Cosmology and Particle Astrophysics, National Taiwan University, No.1, Sec.4, Roosevelt Road, Taipei, 10617 Taipei, Taiwan (R.O.C)
\\
{}$^{2}$ Dipartimento di Fisica e Astronomia ``G. Galilei'', Universit\`{a} degli Studi di
Padova, via Marzolo 8,  I-35131 Padova, Italy
\\
{}$^{3}$ INFN, Sezione di Padova, via Marzolo 8,  I-35131 Padova, Italy
\\
{}$^{4}$ Gran Sasso Science Institute, INFN, Viale F. Crispi 7,  I-67100 L'Aquila, Italy
}

\begin{abstract}

We show that very general scalar-tensor theories of gravity (including, e.g., Horndeski models) are generically invariant under disformal transformations. However there is a special subset, when the transformation is not invertible, that yields new equations of motion which are a generalization of the so-called ``mimetic" dark matter theory recently introduced by
Chamsedinne and Mukhanov. These conclusions hold true irrespective of whether the scalar field in the action of the assumed scalar-tensor theory of gravity is the same or different than the scalar field involved in the transformation.
The new equations of motion for our general mimetic theory can also be derived from an action containing an additional Lagrange multiplier field.
The general mimetic scalar-tensor theory has the same number of derivatives in the equations of motion as the original scalar-tensor theory.
As an application we show that the simplest mimetic scalar-tensor model is able to mimic the cosmological background of a flat FLRW model with a barotropic perfect fluid with any constant equation of state.
\end{abstract}


\date{\today}
\maketitle

\section{Introduction\label{sec:INT}}

The so-called $\Lambda$CDM model of cosmology has been very successful at explaining all cosmological observations with a minimal set of six cosmological parameters. These six parameters have recently been measured to an unprecedented accuracy with the Planck satellite \cite{Adam:2015rua,Ade:2015xua}. Several deviations from this simple model have been constrained to be relatively small \cite{Adam:2015rua,Ade:2015xua,Ade:2015ava,Ade:2015lrj,Ade:2015rim,Ade:2015cva}. This model has been called the standard model of cosmology. However, despite its successes, the $\Lambda$CDM model contains at its core two unknown components. One being a very small cosmological constant and the other the so-called dark matter. This dark matter component is assumed to be cold, collisionless and does not interact with the other particles of the standard model of particle physics except gravitationally. It amounts to about one quarter of the total energy density budget of the current universe \cite{Ade:2015xua}.
Furthermore, despite the many experimental searches for these dark matter particles, both on Earth and in space, none was found up to now. See for example \cite{Garrett:2010hd} and \cite{Bauer:2013ihz} (and reference therein) for reviews.

The presence of these unknown energy components has motivated many studies that try to explain the phenomena that they give origin to, by modifying the law of gravitation without introducing new energy sources, see e.g. \cite{Tsujikawa:2010zza,Clifton:2011jh}.

One recent interesting attempt to explain the dark matter phenomenon by introducing a modification to General Relativity is the so-called ``mimetic" dark matter proposal \cite{Chamseddine:2013kea}. In this proposal, General Relativity is reformulated in terms of an auxiliary metric which is conformally related with the original ``physical" metric, where the conformal factor is a certain function of the new metric and the first derivative of a scalar field. In these new variables, the theory becomes invariant under Weyl rescaling of the new metric and therefore provides traceless equations of motion. This switches on a new conformal degree of freedom of gravity which behaves as an irrotational pressureless perfect fluid, i.e. it can mimic a cold dark matter component.
In this model, the observed cold dark matter energy density would, in general, be the sum of two unknown amounts of energy density contributions, one coming from hypothetical dark matter particles and the other from the ``mimetic" dark matter which is only a gravitational effect.
In a subsequent work \cite{Chamseddine:2014vna} it was shown that by introducing a potential for the new scalar field one can mimic the gravitational behavior of almost any form of matter (see also \cite{Lim:2010yk} for earlier work).
Refs. \cite{Chamseddine:2014vna,Mirzagholi:2014ifa,Capela:2014xta} considered the effects of higher-derivative interactions on the cosmology of the ``mimetic" dark matter model. In \cite{Mirzagholi:2014ifa}, the authors showed that the energy momentum tensor of the mimetic theory with higher-derivatives is actually of the imperfect fluid type and the theory can support vorticity. Ref. \cite{Capela:2014xta} argues that these higher-derivative interactions could possibly help solving the small-scale problems of the cold dark matter model.
Ref. \cite{Barvinsky:2013mea} proposed an alternative conformal extension of General Relativity, using a vector field, that can also support rotational flows for the mimetic dark matter.

The stability of the ``mimetic" dark matter model was analyzed in \cite{Barvinsky:2013mea}, where it was shown that the positiveness of the energy density of the fluid is a sufficient condition for the absence of ghost instabilities. The puzzle of why a simple reparametrization of variables can lead to new additional solutions of the equations of motion was also explained in \cite{Barvinsky:2013mea}.
In \cite{Deruelle:2014zza}, the issue of why new extra solutions are introduced by a reparametrization of variables was revisited from a different viewpoint. The authors showed, in a clear and elegant way, that Einstein's theory of General Relativity is invariant under generic disformal transformations of the type $g_{\mu\nu}=A(\Psi,w)\ell_{\mu\nu}+B(\Psi,w)\partial_\mu\Psi\partial_\nu\Psi$, where $w\equiv\ell^{\mu\nu}\partial_\mu\Psi\partial_\nu\Psi$, $A$ and $B$ are arbitrary functions of their two variables\footnote{We will assume $A>0$. In addition, the transformation should preserve the Lorentzian signature, it must be causal and the transformation for the inverse metric and the volume element should be non-singular. These impose additional conditions on the free functions $A$ and $B$, see \cite{Bettoni:2013diz} for the details.}, $g_{\mu\nu}$ is the ``physical" metric and $\ell_{\mu\nu}$ is the new auxiliary metric. $\Psi$ is a scalar field. This type of disformal transformations were first considered in \cite{Bekenstein:1992pj}. See also for example \cite{Zumalacarregui:2013pma} and references therein. Additionally, they showed that there exists a particular subset of the previous general case, such that the resulting equations of motion are no longer the general relativistic equations but instead one finds the equations of motion of the so-called ``mimetic" dark matter model (also called ``mimetic gravity") \cite{Chamseddine:2013kea}.

In \cite{Golovnev:2013jxa} (see also \cite{Barvinsky:2013mea,Chamseddine:2014vna}), it was shown that the equations of motion of mimetic gravity can be derived by extremizing, with respect to $g_{\mu\nu}$, the Einstein-Hilbert action with the addition of the term $\int d^4x\sqrt{-g}\lambda(g^{\mu\nu}\partial_\mu\Psi\partial_\nu\Psi-1)$, where $\lambda$ is a new field playing the role of a Lagrange multiplier.

The invariance of cosmological perturbations under disformal transformations has been recently studied in for example \cite{Creminelli:2014wna,Minamitsuji:2014waa,Tsujikawa:2014uza,Watanabe:2015uqa,Motohashi:2015pra,Domenech:2015hka} and mimetic theories of modified gravity have been considered in \cite{Myrzakulov:2015qaa} and references therein.

In this paper, we will generalize some results of \cite{Golovnev:2013jxa,Barvinsky:2013mea,Deruelle:2014zza} to a very general scalar-tensor theory of gravity. Our results will be valid for a very general theory, however, for concreteness one may think of the scalar-tensor theory as being the most general healthy second-order theory known as the Horndeski model \cite{Horndeski:1974wa} (see also \cite{Deffayet:2011gz} for a recent rederivation and \cite{Kobayashi:2011nu} for another proof of equivalence with the original formulation of Horndeski). One may also think of it as the recently proposed healthy extensions of the Horndeski model, the so-called $G^3$ theories \cite{Gleyzes:2014dya,Gleyzes:2014qga} or even their extensions \cite{Gao:2014soa}.

This paper is organized as follows. In the next section, we will show under which conditions the previous disformal transformation is non-invertible. Then we will show that very general scalar-tensor theories of gravity are invariant under generic disformal transformations. For a particular special subset of those generic disformal transformation the invariance is broken and one finds new equations of motion which are a generalization of the so-called ``mimetic" dark matter theory. We will show that the invariance is broken exactly for transformations that satisfy the non-invertibility condition.
In section \ref{sec:LM}, we will demonstrate that the new mimetic general scalar-tensor theory equations of motion can also be derived by the use of a Lagrange multiplier as in the General Relativity case. We also briefly comment on the higher-derivative nature of the resulting equations.
In section \ref{sec:EOM}, we shall present some applications of our results. For instance we will show that the simplest mimetic scalar-tensor model is able to mimic the cosmological background of a flat Friedmann-Lema\^{i}tre-Robertson-Walker (FLRW) model with a barotropic perfect fluid with an arbitrary equation of state. Finally, section \ref{sec:CON} is devoted to the conclusions. We use the metric signature $(-,+,+,+)$ and we set the reduced Planck mass to unity.


\section{Mimetic gravity from a disformal transformation\label{sec:DIS}}

In this section, we will consider disformal transformations of very general scalar-tensor theories of gravity. We will show that these theories are invariant under generic disformal transformations. However for a special subset of non-invertible disformal transformations the theory is modified resulting in new equations of motion which may possess novel solutions.

\subsection{Non-invertibility condition of a disformal transformation\label{subsec:NIDIS}}

In this first subsection we will derive what is the condition for non-invertibility of a disformal transformation of the type
\begin{equation}
g_{\mu\nu}=A(\Psi,w)\ell_{\mu\nu}+B(\Psi,w)\partial_\mu\Psi\partial_\nu\Psi,\label{distrans}
\end{equation}
where $w$ is defined as
\begin{equation}
w\equiv \ell^{\rho\sigma}\partial_\rho\Psi\partial_\sigma\Psi.
\end{equation}
$A$ and $B$ are arbitrary functions (see footnote 1) of two variables. $g_{\mu\nu}$ is the original metric and $\ell_{\mu\nu}$ is an auxiliary new metric. $\Psi$ is a scalar field that defines the transformation. In this section, we assume that $\Psi$ is the same scalar field that is present in the action of the scalar-tensor theory. In appendix \ref{GDT} we will consider the case when the disformal transformation introduces a different new field. The issue of the non-invertibility for a conformal transformation in the context of ``mimetic" gravity was first discussed in \cite{Barvinsky:2013mea} and here we will generalize their arguments to disformal transformation in scalar-tensor theories. The conditions under which disformally coupled theories can be re-written in the so-called Jordan frame were studied in \cite{Zumalacarregui:2013pma}.

The inverse of $g_{\mu\nu}$ can be written as
\begin{equation}
g^{\mu\nu}=\frac{1}{A(\Psi,w)}\ell^{\mu\nu}+\frac{B(\Psi,w)}{B(\Psi,w)g^{\rho\sigma}\partial_\rho\Psi\partial_\sigma\Psi-1}g^{\mu\alpha}g^{\nu\beta}\partial_\alpha\Psi\partial_\beta\Psi,\label{invdistrans}
\end{equation}
where $B(\Psi,w)g^{\rho\sigma}\partial_\rho\Psi\partial_\sigma\Psi-1\neq0$ for obvious reasons. Using the previous equation one finds $\ell_{\mu\alpha}\ell^{\alpha\nu}=\delta_\mu^\nu$.

Eq. (\ref{distrans}) is a convolved transformation law for $g_{\mu\nu}$ in terms of $\ell_{\mu\nu}$ because $\ell_{\mu\nu}$ enters in $w$.
In order words, for a fixed $\Psi$ in Eq. (\ref{distrans}), one can see that in order to write $\ell_{\mu\nu}$ is terms of $g_{\mu\nu}$ one needs to solve $w$ in terms of $g_{\mu\nu}$. Here we are assuming that $\Psi$ is not a new variable (i.e. it is already present in the action) because if it was then it would be obvious that the transformation would never in general be invertible. This is because we would be relating ten variables of $g_{\mu\nu}$ to eleven variables of $\ell_{\mu\nu}$ and $\Psi$. Despite this assumption, in the next subsection we will show that the condition we find here for non-invertibility of the transformation is the same as the condition found in the next subsection for the system of equations of motion to be indeterminate. This later condition is valid independently of the assumption that $\Psi$ is a field already present in the action.

Using Eq. (\ref{invdistrans}) one can show that
\begin{equation}
w=\frac{A(\Psi,w)g^{\mu\nu}\partial_\mu\Psi\partial_\nu\Psi}{1-B(\Psi,w)g^{\alpha\beta}\partial_\alpha\Psi\partial_\beta\Psi}.\label{auxw}
\end{equation}
The previous equation can be written as
\begin{equation}
G(\Psi,w)=g^{\mu\nu}\partial_\mu\Psi\partial_\nu\Psi,\label{Geq}
\end{equation}
where the function $G(\Psi,w)$ is defined as
\begin{equation}
G(\Psi,w)\equiv\frac{w\left(1-B(\Psi,w)g^{\alpha\beta}\partial_\alpha\Psi\partial_\beta\Psi\right)}{A(\Psi,w)}.
\end{equation}
For a fixed given $\Psi$ and using the inverse function theorem, if $\frac{dG(\Psi,w)}{dw}|_{w=w_*}\neq0$ then the inverse function $G^{-1}$ exists in a neighborhood of $w_*$ so one can write $w$ as a function of $g_{\mu\nu}$ only as $w=G^{-1}(g^{\mu\nu}\partial_\mu\Psi\partial_\nu\Psi)$. Finally one can use Eq. (\ref{distrans}) (or Eq. (\ref{invdistrans})) to write $\ell_{\mu\nu}$ as a function of $g_{\mu\nu}$. This completes the proof that the inverse transformation, i.e. $\ell_{\mu\nu}(g_{\alpha\beta})$, exists.
Furthermore, the non-existence of $G^{-1}$ implies that $\frac{dG(\Psi,w)}{dw}|_{w=w_*}=0$. One can solve this as
\begin{equation}
G(\Psi,w)=1/b(\Psi),
\end{equation}
where in the right-hand-side we wrote $1/b(\Psi)$ to use the same conventions of notation as in the literature.

If we are in the exceptional case of the previous equation then the transformation from $g_{\mu\nu}$ to $\ell_{\mu\nu}$ cannot be inverted even implicitly and from Eq. (\ref{Geq}) one can find that
\begin{equation}
b(\Psi)=\frac{1}{g^{\mu\nu}\partial_\mu\Psi\partial_\nu\Psi}.\label{normalizationcondition}
\end{equation}
The previous equation can be used with Eq. (\ref{auxw}) to find
\begin{equation}
B(\Psi,w)=-\frac{A(\Psi,w)}{w}+b(\Psi).\label{mimeticdis}
\end{equation}
This condition for having a non-invertible transformation is the same as the condition that Deruelle and Rua \cite{Deruelle:2014zza} found for the system of equations of motion of mimetic dark matter to be indeterminate. We will generalize their analysis in the next subsection. Note that here we never assumed any explicit scalar-tensor theory so this result is very general. Eq. (\ref{normalizationcondition}) is a kinematical constraint valid independently of the dynamics. Furthermore, the results of this subsection also explain why it is not so surprising that the transformed scalar-tensor theory, i.e. mimetic gravity, may contain new solutions with respect to the original theory. The reason is that we are performing a non-invertible change of variables.

\subsection{Disformal transformation method \label{subsec:DISM}}

In this subsection, we will perform a disformal transformation of the type (\ref{distrans}) on a very general scalar-tensor theory and compute the equations of motion that result. This is a generalization of the results in Deruelle and Rua \cite{Deruelle:2014zza}. The further generalization for the case when the transformation field is different from the scalar field in the action is discussed in Appendix  \ref{GDT}. This is the case in the ``mimetic" dark matter model \cite{Chamseddine:2013kea}.

We start with a very general local action of the type
\begin{eqnarray}
S&=&\int d^4x\sqrt{-g}\mathcal{L}[g_{\mu\nu},\partial_{\lambda_1}g_{\mu\nu},\ldots,\partial_{\lambda_1}\ldots\partial_{\lambda_p}g_{\mu\nu},\Psi,\partial_{\lambda_1}\Psi,\ldots,\partial_{\lambda_1}\ldots\partial_{\lambda_q}\Psi]+S_m[g_{\mu\nu},\phi_m]
,\label{STgeneralaction}
\end{eqnarray}
where the integers $p,q\geq 2$, $\mathcal{L}$ is the Lagrangian density which is a functional of the metric, the scalar field and their derivatives. $S_m$ is the action for the matter field $\phi_m$ which we assume to be uncoupled with $\Psi$. For the sake of concreteness, the Lagrangian $\mathcal{L}$ may be thought of as being the Lagrangian of Horndeski's theory \cite{Horndeski:1974wa} or one of its recently proposed healthy extensions \cite{Gleyzes:2014dya,Gleyzes:2014qga,Gao:2014soa}.

The variation of the action with respect to the fundamental fields, $\Psi$, $g_{\mu\nu}$ and $\phi_m$, is given by,
\begin{equation}
\delta S={1\over2}\int\!d^4x\sqrt{-g}(E^{\mu\nu}+T^{\mu\nu})\delta g_{\mu\nu} + \int\!d^4x\ \Omega_\Psi \delta \Psi + \int\!d^4x\ \Omega_m \delta \phi_m,\label{old_S_variation}
\end{equation}
where
\begin{eqnarray}
\Omega_\Psi&=&\frac{\delta\left(\sqrt{-g}\mathcal{L}\right)}{\delta\Psi}=\frac{\partial(\sqrt{-g}\mathcal{L})}{\partial\Psi}+\sum_{h=1}^q(-1)^h\frac{d}{dx^{\lambda_1}}\ldots\frac{d}{dx^{\lambda_h}}\frac{\partial(\sqrt{-g}\mathcal{L})}{\partial\left(\partial_{\lambda_1}\ldots\partial_{\lambda_h}\Psi\right)},
\\
E^{\mu\nu}&=&\frac{2}{\sqrt{-g}}\frac{\delta(\sqrt{-g}\mathcal{L})}{\delta g_{\mu\nu}}=\frac{2}{\sqrt{-g}}\left(\frac{\partial(\sqrt{-g}\mathcal{L})}{\partial g_{\mu\nu}}+\sum_{h=1}^p(-1)^h\frac{d}{dx^{\lambda_1}}\ldots\frac{d}{dx^{\lambda_h}}\frac{\partial(\sqrt{-g}\mathcal{L})}{\partial\left(\partial_{\lambda_1}\ldots\partial_{\lambda_h}g_{\mu\nu}\right)}\right),
\\
T^{\mu\nu}&=&\frac{2}{\sqrt{-g}}\frac{\delta (\sqrt{-g}\mathcal{L}_m)}{\delta g_{\mu\nu}}, \,
\Omega_m=\frac{\delta (\sqrt{-g}\mathcal{L}_m)}{\delta \phi_m},
\,\mathrm{where}\,\, S_m[g_{\mu\nu}]=\int d^4x\sqrt{-g}\mathcal{L}_m[g_{\mu\nu},\phi_m].
\end{eqnarray}
$\mathcal{L}_m$ is the matter Lagrangian density and $T^{\mu\nu}$ is the matter energy-momentum tensor. In the case of General Relativity, the tensor $E^{\mu\nu}$ is $E^{\mu\nu}=-G^{\mu\nu}$, where $G^{\mu\nu}$ is the Einstein tensor.

We consider a disformal transformation of the type (\ref{distrans}) from where one can obtain its variation as
\begin{eqnarray}
\delta g_{\mu\nu}&=&A\,\delta\ell_{\mu\nu}-\left(\ell_{\mu\nu}{\partial A\over\partial w}+\partial_\mu\Psi\partial_\nu\Psi{\partial B\over \partial w}\right)\left[(\ell^{\alpha\rho}\partial_\alpha\Psi)\,
(\ell^{\beta\sigma}\partial_\beta\Psi)\,\delta\ell_{\rho\sigma}-2\ell^{\rho\sigma}(\partial_\rho\Psi)\,(\partial_\sigma\delta\Psi)\right]\nonumber\\
&&+\left(\ell_{\mu\nu}{\partial A\over\partial  \Psi}+\partial_\mu\Psi\partial_\nu\Psi{\partial B\over \partial \Psi}\right)\delta\Psi +B\left[(\partial_\mu\Psi)(\partial_\nu\delta\Psi)+(\partial_\nu\Psi)(\partial_\mu\delta\Psi)\right]. \label{g_variation}
\end{eqnarray}
Inserting Eq. (\ref{g_variation}) into Eq. (\ref{old_S_variation}), the generalized Einstein equations of motion, $\delta S/\delta\ell_{\mu\nu}=0$, are
\begin{eqnarray}
 A(E^{\mu\nu}+T^{\mu\nu})&=&\left(\alpha_1{\partial A\over\partial w}+\alpha_2{\partial B\over\partial w}\right)(\ell^{\mu\rho}\partial_\rho\Psi)\,
(\ell^{\nu\sigma}\partial_\sigma\Psi), \label{new_E_eom}
\end{eqnarray}
and the generalized Klein-Gordon equation, $\delta S/\delta\Psi=0$, is,
\begin{equation}
{1\over\sqrt{-g}}\partial_\rho\left\{\sqrt{-g}\,\partial_\sigma\Psi\left[B(E^{\rho\sigma}+ T^{\rho\sigma})+\left(\alpha_1{\partial A\over\partial w}+\alpha_2{\partial B\over\partial w}\right)\ell^{\rho\sigma}\right]
\right\} - \frac{\Omega_\Psi}{\sqrt{-g}}=
{1\over2}\left(\alpha_1{\partial A\over\partial\Psi}+\alpha_2{\partial B\over\partial\Psi}\right),\label{new_psi_eom}
\end{equation}
where we have defined two new quantities as
\begin{equation}
 \alpha_1 \equiv (E^{\rho\sigma}+ T^{\rho\sigma})\ell_{\rho\sigma}\quad\hbox{and}\quad \alpha_2\equiv(E^{\rho\sigma}+ T^{\rho\sigma})\partial_\rho\Psi\,\partial_\sigma\Psi.
\end{equation}
In addition, the equation of motion for the matter field is $\Omega_m=0$.

By contracting the metric equations of motion (\ref{new_E_eom}) with $\ell_{\mu\nu}$  and  with $\partial_\mu\Psi\partial_\nu\Psi$, we find
\begin{equation}
\alpha_1\left(A-w{\partial A\over\partial w}\right)-\alpha_2 w{\partial B\over\partial w}=0,\quad\quad \alpha_1\,w^2{\partial A\over\partial w}-\alpha_2\left(A-w^2{\partial B\over\partial w}\right)=0.\label{system_eqn}
\end{equation}
These two equations form a two-dimensional linear system of algebraic equations for $\alpha_1$ and $\alpha_2$. The solutions of the system are different depending on whether its determinant is zero or non-zero. In the next two subsections we study these two cases separately.

\subsubsection{Generic case \label{subsubsec:generic}}

We may write the system of equations (\ref{system_eqn}) in matrix form, as
\begin{eqnarray}
M
\begin{pmatrix}
\alpha_1\\\alpha_2
\end{pmatrix}
=0
,\qquad\mathrm{where}\qquad M =
 \begin{pmatrix}
A-w{\partial A\over\partial w} & - w{\partial B\over\partial w}\\
&&\\
w^2{\partial A\over\partial w} & -A+w^2{\partial B\over\partial w}
 \end{pmatrix}.
\end{eqnarray}
The determinant of the system is
\begin{equation}
\mathrm{det} (M) =w^2A{\partial\ \over\partial w}\left(B+{A\over w}\right).\label{det}
\end{equation}
If $\mathrm{det} (M) \neq 0$ then the only solution is $\alpha_1=\alpha_2=0$. For this generic case the equations of motion, Eqs. (\ref{new_E_eom}) and (\ref{new_psi_eom}), reduce to
\begin{eqnarray}
E^{\mu\nu}+T^{\mu\nu}&=&0,\\
\Omega_\Psi = 0. \label{recover_KG_eqn_1}
\end{eqnarray}
When written in terms of the metric $g_{\mu\nu}$, these two equations in addition to $\Omega_m=0$ are the same equations as in the original theory before doing any disformal transformation. In other words, by taking the variation with respect to the original metric $g_{\mu\nu}$ or with respect to $\ell_{\mu\nu}$ and $\Psi$ we get, in the end, the same equations of motion. This shows that generically (i.e. $\mathrm{det} (M) \neq 0$) the theory is invariant under disformal transformations of the type (\ref{distrans}). This generalizes the results of \cite{Deruelle:2014zza} (obtained for Einstein gravity) to a very general scalar-tensor theory of the kind (\ref{STgeneralaction}). This result is less surprising if one recalls that all one is doing is a well-behaved invertible change of variables.

\subsubsection{Mimetic gravity \label{subsubsec:MG}}

If the determinant of the system is zero then one can solve the differential equation (\ref{det}) to find that the free function $B(\Psi,w)$ has to be of the form
\begin{equation}
B(\Psi, w)= -{A(\Psi,w)\over w}+b(\Psi),\label{mimetic B}
\end{equation}
where $b(\Psi)$ is an integration constant (it does not depend on $w$ but it may depends on $\Psi$) and we assume it is non-zero for all $\Psi$.
This solution was previously found in \cite{Deruelle:2014zza} for the case when the starting action in Eq. (\ref{STgeneralaction}) is simply the Einstein-Hilbert action. Here we show that solution (\ref{mimetic B}) is still valid for a general action of the form (\ref{STgeneralaction}), irrespective of whether the scalar field in the action is the same or different than the scalar field involved in the transformation, as shown in Appendix  \ref{GDT}. This is a consequence of the fact  that the determinant of the system, Eq. (\ref{det}), does not depend on the form of the starting action (\ref{STgeneralaction}) and it is the same as the determinant found in \cite{Deruelle:2014zza}. Substituting solution (\ref{mimetic B}) into the system (\ref{system_eqn}) gives us $\alpha_2=w\alpha_1$. Hence, the equations of motion (\ref{new_E_eom}) and (\ref{new_psi_eom}) become
\begin{equation}
E^{\mu\nu}+T^{\mu\nu}={\alpha_1\over w}\,(\ell^{\mu\rho}\partial_\rho\Psi)\,
(\ell^{\nu\sigma}\partial_\sigma\Psi),\quad\quad
{1\over\sqrt{-g}}\partial_\rho\left(\sqrt{-g}\,b\,\alpha_1\,\ell^{\rho\sigma}\partial_\sigma\Psi\right) - \frac{\Omega_\Psi}{\sqrt{-g}} = {1\over2} \alpha_1\,w{db\over d\Psi}.\label{mixed eom}
\end{equation}
Now, the disformal transformation is of the particular type
\begin{equation}
g_{\mu\nu}=A(\Psi,w)\,\ell_{\mu\nu}+\partial_\mu\Psi\,\partial_\nu\Psi\left(b(\Psi)-{A(\Psi,w)\over w}\right).\label{mimeticdisformaltransformation}
\end{equation}
The inverse metric transforms (recall we assume  $b\neq0$) as
\begin{equation}
g^{\mu\nu}={\ell^{\mu\nu}\over A}+{A-w\,b\over A\,b\,w^2}(\ell^{\mu\rho}\partial_\rho\Psi)\,
(\ell^{\nu\sigma}\partial_\sigma\Psi),
\end{equation}
and these equations can be used to write (\ref{mixed eom}) in terms of $g_{\mu\nu}$ (explicitly) only.
Similarly to \cite{Deruelle:2014zza}, we have  $\ell^{\mu\rho}\partial_\rho\Psi=bw\partial^\mu\Psi$ and $\alpha_1=(E+T)/(bw)$ where $\partial^\mu\Psi\equiv g^{\mu\rho}\partial_\rho\Psi$ and $E+T\equiv g_{\rho\sigma}(E^{\rho\sigma}+ T^{\rho\sigma})$. By contracting $\ell^{\mu\rho}\partial_\rho\Psi=bw\partial^\mu\Psi$ with $\partial_\mu\Psi$ and using the definition of $w$ one can also find that
\begin{equation}
b(\Psi)g^{\mu\nu}\partial_\mu\Psi\partial_\nu\Psi=1.\label{kinematicalc}
\end{equation}
So the equations of motion (\ref{mixed eom}) simplify to
\begin{equation}
E_{\mu\nu}+ T_{\mu\nu}=(E+T)\,b\,\partial_\mu\Psi\,\partial_\nu\Psi,\quad\quad
\nabla_\rho\left[(E+T)b\,\partial^\rho\Psi\right] - \frac{\Omega_\Psi}{\sqrt{-g}} = \frac{1}{2}(E+T) {1\over b}{db\over d\Psi}, \label{EOM_mimetic1}
\end{equation}
where $\nabla_\rho$ denotes the covariant derivative with respect to $g_{\mu\nu}$.
In order to have the full system of equations of motion, to these equations one should add the matter equation, $\Omega_m=0$. As it can be seen these equations of motion are in general different from the equations of motion that result from varying the action (\ref{STgeneralaction}) with respect to the original metric $g_{\mu\nu}$. We will call this new theory ``mimetic" gravity and the transformation will be called mimetic disformal transformation. Note that the condition for the determinant of the system to be zero leads to exactly the same particular disformal transformation, Eq. (\ref{mimeticdisformaltransformation}), as the non-invertibility condition of the previous subsection.

\section{Mimetic gravity from a Lagrange multiplier\label{sec:LM}}

In this section, we will show that the mimetic equations of motion that result after transforming the theory (\ref{STgeneralaction}) via a mimetic disformal transformation (\ref{mimeticdisformaltransformation}) can also be obtained by variation of an action without performing any disformal transformation or introducing an auxiliary metric $\ell_{\mu\nu}$. For the case in which the original theory (\ref{STgeneralaction}) is General Relativity and for conformal transformations this was first achieved in \cite{Golovnev:2013jxa,Barvinsky:2013mea}.

Let us start with the very general action of the previous section where we add an additional term as
\begin{eqnarray}
S_\lambda&=&\int d^4x\sqrt{-g}\mathcal{L}[g_{\mu\nu},\partial_{\lambda_1}g_{\mu\nu},\ldots,\partial_{\lambda_1}\ldots\partial_{\lambda_p}g_{\mu\nu},\Psi,\partial_{\lambda_1}\Psi,\ldots,\partial_{\lambda_1}\ldots\partial_{\lambda_q}\Psi]+S_m[g_{\mu\nu},\phi_m]\nonumber\\
&&+\int d^4x\sqrt{-g}\lambda\left(b(\Psi)g^{\mu\nu}\partial_\mu\Psi\partial_\nu\Psi-1\right)
,\label{generalaction}
\end{eqnarray}
where $\lambda$ is a Lagrange multiplier field which enforces the kinematical constraint. $b(\Psi)$ is a known potential function that defines the theory.
The equations of motion that result from varying the action with respect to $\lambda$, $\Psi$, $g_{\mu\nu}$ and $\phi_m$ are respectively (after some simplification)
\begin{eqnarray}
&&
b(\Psi)g^{\mu\nu}\partial_\mu\Psi\partial_\nu\Psi-1=0,\label{normalization}\\
&&
\Omega_\Psi+\sqrt{-g}\frac{\lambda}{b(\Psi)}\frac{db(\Psi)}{d\Psi}-2\partial_\mu\left(\sqrt{-g}\lambda b(\Psi)g^{\mu\nu}\partial_\nu\Psi\right)=0,\\
&&
E^{\mu\nu}+T^{\mu\nu}-2\lambda b(\Psi)\partial^\mu\Psi\partial^\nu\Psi=0,\label{mEE}\\
&&
\Omega_m=0,
\end{eqnarray}
with the same definitions of Section \ref{sec:DIS}.
Taking the trace of Eq. (\ref{mEE}) and after using Eq. (\ref{normalization}), one obtains
\begin{equation}
2\lambda=E+T,\label{sollambda}
\end{equation}
where $E=g_{\mu\nu}E^{\mu\nu}$ and $T=g_{\mu\nu}T^{\mu\nu}$. One can see that the Lagrange multiplier is given by the traces $E$ and $T$ and this can be used to eliminate $\lambda$ from the equations of motion to obtain
\begin{eqnarray}
&&
b(\Psi)g^{\mu\nu}\partial_\mu\Psi\partial_\nu\Psi-1=0,\label{normalization2}\\
&&
\nabla_\mu\left[(E+T)b(\Psi)\partial^\mu\Psi\right]-\frac{\Omega_\Psi}{\sqrt{-g}}=\frac{E+T}{2}\frac{1}{b(\Psi)}\frac{db(\Psi)}{d\Psi},\label{mKG}\\
&&
E^{\mu\nu}+T^{\mu\nu}=(E+T)b(\Psi)\partial^\mu\Psi\partial^\nu\Psi,\label{mEE2}\\
&&
\Omega_m=0\label{mMF}.
\end{eqnarray}
These equations of motion are the same as the mimetic equations of motion in subsection \ref{subsec:DISM}, i.e. (\ref{kinematicalc}), (\ref{EOM_mimetic1}) and the matter equation. This shows that mimetic gravity can be formulated by action (\ref{generalaction}). The price to pay in this formulation is that one needs to introduce an additional scalar field, the Lagrange multiplier $\lambda$. It would be interesting to determine if it is possible to derive mimetic gravity from an action with no additional scalar fields like $\lambda$. We leave this for future work.

Let us take the covariant derivative of Eq. (\ref{mEE2}) and use $\nabla_\mu T^{\mu\nu}=0$\footnote{The conservation of the energy-momentum tensor is a consequence of assuming that the action $S_m$ can be written as a functional of the matter field and the metric $g_{\mu\nu}$ and by using the Horndeski identity, Eq. (\ref{HorndeskiIdentity}), applied to the matter action together with the equation of motion (\ref{mMF}).} to obtain
\begin{eqnarray}
\nabla_\mu E^{\mu\nu}&=&\nabla_\mu\left[(E+T)b(\Psi)\partial^\mu\Psi\right]\partial^\nu\Psi+(E+T)b(\Psi)\partial^\mu\Psi\nabla_\mu\partial^\nu\Psi\nonumber\\
&=&\partial^\nu\Psi\left[\nabla_\nu\left[(E+T)b(\Psi)\partial^\mu\Psi\right]-\frac{1}{2}\frac{1}{b(\Psi)}\frac{db(\Psi)}{d\Psi}(E+T)\right],\label{auxH}
\end{eqnarray}
where in the second line we have used that $b(\Psi)\partial^\mu\Psi\partial_\mu\Psi=1$ and that from its covariant derivative one obtains $b(\Psi)\nabla^\mu\nabla^\nu\Psi\partial_\mu\Psi=-\frac{1}{2}\frac{db(\Psi)}{d\Psi}\nabla^\nu\Psi\partial_\mu\Psi\partial^\mu\Psi$.
It was shown by Horndeski  \cite{Horndeski:1974wa} (see also references therein) that
\begin{equation}
\sqrt{-g}\nabla_\mu E^{\mu\nu}=\Omega_\Psi\nabla^\nu\Psi.\label{HorndeskiIdentity}
\end{equation}
Using this and the fact that $\partial^\nu\Psi\neq0$ at least for one index $\nu$ we can simplify Eq. (\ref{auxH}) to
\begin{equation}
\nabla_\mu\left[(E+T)b(\Psi)\partial^\mu\Psi\right]-\frac{\Omega_\Psi}{\sqrt{-g}}=\frac{E+T}{2}\frac{1}{b(\Psi)}\frac{db(\Psi)}{d\Psi}.
\end{equation}
This is exactly the same equation as (\ref{mKG}). So we have managed to show that Eq. (\ref{mKG}) results from taking the covariant derivative of  Eq. (\ref{mEE2}) and use $\nabla_\mu T^{\mu\nu}=0$ and Eqs. (\ref{normalization2}) and (\ref{mMF}). This proof is independent of using the Lagrange multiplier method or not and shows that in order to solve the dynamics of the system it is sufficient to consider Eqs. (\ref{normalization2}), (\ref{mEE2}) and (\ref{mMF}).
These three equations, when written in terms of the metric $g_{\mu\nu}$, do not contain any more higher-order derivatives that the equations of motion that result from the non-mimetic theory defined by the Lagrangian $\mathcal{L}$. It is also worth noting that the action (\ref{generalaction}) does not contain any more higher-order derivatives than $\mathcal{L}$. However, the new theory (\ref{generalaction}) does contain a new field, the Lagrange multiplier $\lambda$. The three independent equations of motion when written in terms of the new metric $\ell_{\mu\nu}$ may contain higher-order derivatives.
For concreteness, we can think of $\mathcal{L}$ as being the Horndeski Lagrangian \cite{Horndeski:1974wa}\footnote{One can also consider healthy extensions of Horndeski's theory, like for instance the so-called $G^3$ theories \cite{Gleyzes:2014dya,Gleyzes:2014qga} or even their extensions \cite{Gao:2014soa}.} and we would be considering the ``mimetic" Horndeski theory.

\section{Non-trivial examples of cosmology in the ``mimetic" Horndeski model\label{sec:EOM}}

As an application of the results of the preceding sections, in this section, we will present three simple examples of non-trivial cosmological solutions that arise in very simple ``mimetic" Horndeski models.

The most general class of 4D local scalar-tensor theories that contain second-order equations of motion and that can be derived from an action is known as the Horndeski theory \cite{Horndeski:1974wa}. Its action is
\begin{equation}
S_H=\int d^4x\sqrt{-g}\mathcal{L}_H=\int d^4x\sqrt{-g}\sum_{n=0}^{3}\mathcal{L}_{n},
\label{action}
\end{equation}
where
\begin{eqnarray}
\mathcal{L}_0 & = & K\left(X,\Psi\right),\\
\mathcal{L}_1 & = & -G_3\left(X,\Psi\right)\Box\Psi,\\
\mathcal{L}_2 & = & G_{4,X}\left(X,\Psi\right)\left[\left(\Box\Psi\right)^{2}-\left(\nabla_{\mu}\nabla_{\nu}\Psi\right)^{2}\right]
+R\,G_4\left(X,\Psi\right) ,\\
\mathcal{L}_3 & = & -\frac{1}{6}G_{5,X}\left(X,\Psi\right)\left[\left(\Box\Psi\right)^{3}-3\Box\Psi\left(\nabla_{\mu}\nabla_{\nu}\Psi\right)^{2}
+2\left(\nabla_{\mu}\nabla_{\nu}\Psi\right)^{3}\right]+G_{\mu\nu}\nabla^{\mu}\nabla^{\nu}\Psi\,
G_5\left(X,\Psi\right),
\end{eqnarray}
and $X=-1/2\nabla_\mu\Psi\nabla^\mu\Psi$, $(\nabla_\mu\nabla_\nu\Psi)^2=\nabla_\mu\nabla_\nu\Psi\nabla^\mu\nabla^\nu\Psi$ and $(\nabla_\mu\nabla_\nu\Psi)^3=\nabla_\mu\nabla_\nu\Psi\nabla^\mu\nabla^\rho\Psi\nabla^\nu\nabla_\rho\Psi$. The functions $K(X,\Psi)$, $G_3(X,\Psi)$, $G_4(X,\Psi)$ and $G_5(X,\Psi)$ are free function of their two variables and define a particular theory in the Horndeski class. The subscript ${}_{,X}$ denotes derivative with respect to $X$.
In the next three subsections the actions of the models considered will be Eq. (\ref{generalaction}) with $\mathcal{L}=\mathcal{L}_H$ but with different choices for the free functions in each subsection.

Notice that for a general mimetic Horndeski model, the free function $b(\Psi)$ in the second term of Eq. (\ref{generalaction}) can be reabsorbed by defining a new field $\Phi$ as $d\Phi=\sqrt{|b|}d\Psi$. Because the Horndeski Lagrangian is form invariant under field redefinitions of this type, this transformation just amounts to consider a different starting Horndeski Lagrangian $\mathcal{L}_H$.

\subsection{A very simple example \label{subsec:SE}}

In our first simple example we will consider the mimetic theory of a canonical kinetic term scalar field with no potential coupled to Einstein's gravity theory. The action of this model is Eq. (\ref{generalaction}) with $\mathcal{L}=\mathcal{L}_H$, $S_m=0$ and with the choice
\begin{eqnarray}
K(X,\Psi)=c_2X,\qquad G_3(X,\Psi)=0, \qquad G_4(X,\Psi)=1/2,\qquad G_5(X,\Psi)=0,
\end{eqnarray}
where $c_2$ is a constant which may have either sign. In the non-mimetic theory, if $c_2$ is negative it is well known that the scalar field has the wrong sign in the kinetic term and is a ghost. In the present mimetic model it would be interesting to study perturbations and determine what are the conditions for the absence of ghost and other instabilities. We leave this work for the near future.
With the same notation of the previous section and for a flat FLRW background, the objects that appear there are
\begin{eqnarray}
E_{00}&=&-3 H^2 + \frac{1}{2} c_2 \dot\Psi^2,\\
E_{xx}=E_{yy}=E_{zz}&=&a^2 (3 H^2 +
   2 \dot H + \frac{1}{2} c_2 \dot\Psi^2 ),\\
E&=&12 H^2 + 6 \dot H + c_2 \dot\Psi^2,\\
\Omega_\Psi&=&a^3 \left(-3 c_2 H \dot\Psi
 - c_2 \ddot\Psi
\right),
\end{eqnarray}
where $a$ is the scale factor, $H=\dot a/a$ and dot denotes derivative with respect to cosmic time. ${x,y,z}$ denote the comoving spatial coordinates.
The equations of motion for this simple mimetic model, i.e. Eq. (\ref{normalization2}), the time and spatial components of Eq. (\ref{mEE2}) and Eq. (\ref{mKG}), are respectively
\begin{eqnarray}
&&\!\!\!\!\!\!\!\!
b(\Psi)\dot\Psi^2+1=0,\label{nor}\\
&&\!\!\!\!\!\!\!\!
3H^2=\frac{\dot\Psi^2}{2}\left[c_2-2b(\Psi)\left(12H^2+6\dot H+c_2\dot\Psi^2\right)\right],\label{hamil}\\
&&\!\!\!\!\!\!\!\!
6H^2+4\dot H+c_2\dot\Psi^2=0,\label{FriedmannEq}\\
&&\!\!\!\!\!\!\!\!
b(\Psi)\left[-6H(6H^2\dot\Psi+7\dot\Psi\dot H+2H\ddot\Psi)-6\dot H\ddot\Psi-6\dot\Psi\ddot H-3c_2\dot\Psi^2(H\dot\Psi+\ddot\Psi)\right]
\nonumber\\
&&\quad\,\,
+c_2 (3H\dot\Psi+\ddot\Psi)+b'(\Psi)\left(\frac{1}{b(\Psi)}+2\dot\Psi^2\right)\left[-6H^2-3\dot H-\frac{c_2}{2}\dot\Psi^2\right]=0,\label{kmaux}
\end{eqnarray}
where prime denotes derivative with respect to the field $\Psi$. Eqs. (\ref{hamil}) and (\ref{kmaux}) are not independent from Eqs. (\ref{nor}) and (\ref{FriedmannEq}) because they can be derived from them.

It is easy to check that Eqs. (\ref{nor}) and (\ref{FriedmannEq}) admit the following solution
\begin{eqnarray}
a(t)=t^\frac{2}{3(1+\omega)}, \quad \Psi(t)=\pm\sqrt{-\frac{\alpha}{c_2}}\log\frac{t}{t_0},\quad b(\Psi)=-\frac{1}{\dot\Psi^2}=\frac{c_2}{\alpha}t^2=\frac{c_2}{\alpha} t_0^2 e^{\pm2\sqrt{-\frac{c_2}{\alpha}}\Psi},
\end{eqnarray}
where $t_0$ is an integration constant, the parameter $\alpha$ is $\alpha=-\frac{8\omega}{3(1+\omega)^2}$, where $\omega$ is a constant parameter. This expansion law is the same as the one given by a perfect fluid universe with a constant equation of state $\omega$. If $c_2$ is positive then the equation of state $\omega$ has to be positive too. This shows that this simple mimetic scalar field model can mimic the background evolution of a perfect fluid universe with a constant equation of state. For different $\omega$ the value of $\alpha$ changes but the functional form of $b(\Psi)$ does not change. It is obvious that this new solution is not a solution of the Einstein plus Klein-Gordon (with zero potential) field theory. There $\omega$ is necessarily unity.

By adjusting the function $b(\Psi)$ accordingly (note that $b(\Psi)<0$ for a time-like scalar velocity), this simple model can mimic the expansion history of almost any model. To be concrete, we can mimic the expansion history of a perfect fluid model with a fixed sign of the pressure. In that case, $6H^2+4\dot H=-2p$, where $p$ is the pressure of the perfect fluid. So from the independent equation of motion (\ref{FriedmannEq}) one can see that the pressure cannot change sign. The fact that one can have almost any expansion history desired is somewhat similar to the minimal extension of the original mimetic dark matter model proposed in \cite{Chamseddine:2014vna}. See also \cite{Lim:2010yk} for an earlier work where models similar to our present one were considered.

\subsection{Mimetic cubic Galileon \label{subsec:MCG}}

In this subsection, we will consider the mimetic cubic Galileon model as a further example of a simple mimetic Horndeski model. The mimetic cubic Galileon model with $c_2=0$ (and also including the other Galileon interactions) was previously studied in \cite{Haghani:2015iva} but for the case of a constant $b(\Psi)$. Here we allow the function $b$ to depend on $\Psi$.
The action of the model is Eq. (\ref{generalaction}) with $\mathcal{L}=\mathcal{L}_H$, $S_m=0$ and with the free functions chosen as
\begin{eqnarray}
K(X,\Psi)=c_2X,\qquad G_3(X,\Psi)=2c_3/\tilde{\Lambda}^3X, \qquad G_4(X,\Psi)=1/2,\qquad G_5(X,\Psi)=0,
\end{eqnarray}
where from now on we will set the cutoff scale $\tilde{\Lambda}$ to be $\tilde{\Lambda}=1$ and $c_3$ is a new model parameter.

\begin{figure}[h]
\includegraphics{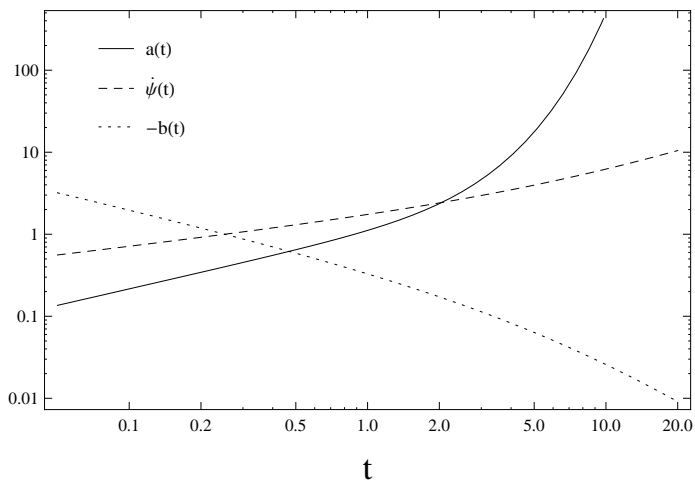}
\caption{Plot of the scale factor $a(t)$ (solid line), the time derivative of field $\dot\Psi(t)$ (dashed line) and the function $-b(t)$ (dotted line) as functions of time $t$ (in suitable units) for the parameter choice $\mathcal{C}=c_2=c_3=a_\star=1$. This choice was made for illustration purposes only.}
\label{Fig:plot1}
\end{figure}

Analogously to the previous subsection, there are only two independent equations of motion, they can be chosen to be Eq. (\ref{normalization2}) and the spatial component of Eq. (\ref{mEE2}). They are respectively
\begin{eqnarray}
&&\!\!\!\!\!\!\!\!
b(\Psi)\dot\Psi^2+1=0,\label{normalizationcubic}\\
&&\!\!\!\!\!\!\!\!
6H^2+4\dot H+\dot\Psi^2(c_2-4c_3\ddot\Psi)=0.\label{FriedmannEqcubic}
\end{eqnarray}
As in the preceding section by suitably choosing a function $b(\Psi)$ one can have almost any expansion history desired. Let us for instance concentrate on the expansion history of a universe filled with dark matter and a positive cosmological constant $\Lambda$. The scale factor solution for that universe is
\begin{equation}
a=a_\star\sinh^\frac{2}{3}(\mathcal{C}t),
\end{equation}
where $\mathcal{C}=\sqrt{3\Lambda/4}$.
Eq. (\ref{FriedmannEqcubic}) can be integrated once to find
\begin{equation}
\frac{4c_3}{c_2}\left[-\arctan\left(\pm\sqrt{\frac{3c_2}{8\mathcal{C}^2}}\dot\Psi\right)\pm\sqrt{\frac{3c_2}{8\mathcal{C}^2}}\dot\Psi\right]=t.
\end{equation}
In Fig. \ref{Fig:plot1} we plot the time evolution of the scale factor $a(t)$, the time derivative of $\Psi$ and by using Eq. (\ref{normalizationcubic}) one can find the function $b(t)$. For illustration purposes we choose the model parameters as $\mathcal{C}=c_2=c_3=a_\star=1$. For this choice, the matter-dominated era ends around $t=\mathcal{O}(1)$ and after that the universe becomes dominated by the energy density of the cosmological constant. For $\mathcal{C}t\gg1$, the time derivative of $\Psi$ is $\dot\Psi\propto t$ while for $\mathcal{C}t\ll1$ it becomes $\dot\Psi\propto t^{1/3}$. The previous equations can be easily integrated to find the function $b(\Psi)$ as $b(\Psi)\propto-\Psi^{-1/2}$ for $\mathcal{C}t\ll1$ and $b(\Psi)\propto-\Psi^{-1}$ for $\mathcal{C}t\gg1$. By choosing a function $b(\Psi)$ with these asymptotic limits one can approximately reproduce the expansion history of a $\Lambda$-dark matter universe.

\subsection{The case of minimal coupling to $\ell_{\mu\nu}$ \label{subsec:ellmunu}}

The third example of non-trivial cosmological solutions that arise in the context of mimetic Horndeski models that we are going to present now involves promoting the auxiliary metric $\ell_{\mu\nu}$ to the physical metric. Say for instance, usual matter, like baryons, are minimally coupled with $\ell_{\mu\nu}$ instead of the more interesting case of minimal coupling with $g_{\mu\nu}$. The gravitational part of the action of this model is Eq. (\ref{STgeneralaction}) with $S_m=0$, where the fundamental metric variable is the metric $\ell_{\mu\nu}$, which is related to the metric $g_{\mu\nu}$ by a mimetic disformal transformation, i.e. a disformal transformation of the type (\ref{distrans}) with the function $B$ given by (\ref{mimeticdis}). Then we choose to minimally couple this gravitational theory for $\ell_{\mu\nu}$ and $\Psi$ with (baryon) matter fields.

In the following discussion, we will restrict the mimetic disformal transformation to a particular type (see Eq. (\ref{pmdt2}) below) so that we have a Weyl symmetry, that is, the gravitational part of the action will be invariant under a Weyl rescaling of the type $\ell_{\mu\nu}\rightarrow\Omega^2(x)\ell_{\mu\nu}$, where $\Omega(x)$ is a non-zero function. It is worth mentioning that in the case of a mimetic disformal transformation with $B(\Psi,w)=0$ (the disformal transformation becomes a conformal transformation) this implies that $A(\Psi,w)=b(\Psi)w$. It is then easy to see that the theory also has a Weyl symmetry which allows us to choose the gauge so that $g_{\mu\nu}=\ell_{\mu\nu}$ as it was done in \cite{Barvinsky:2013mea}. Indeed, in \cite{Chamseddine:2013kea,Chamseddine:2014vna,Barvinsky:2013mea}, the authors used $B(\Psi,w)=0$, $b(\Psi)=-1$ which leads to $A(\Psi,w)= - w$.

If the function $A(\Psi,w)$ is
\begin{equation}
A(\Psi,w)=(b(\Psi)-f(\Psi))w,\label{pmdt1}
\end{equation}
then the mimetic disformal transformation is
\begin{equation}
g_{\mu\nu}=(b(\Psi)-f(\Psi))w\ell_{\mu\nu}+f(\Psi)\partial_\mu\Psi\partial_\nu\Psi,\label{pmdt2}
\end{equation}
and the inverse metric transformation is
\begin{equation}
g^{\mu\nu}=\frac{\ell^{\mu\nu}}{(b(\Psi)-f(\Psi))w}-\frac{f(\Psi)}{w^2b(\Psi)(b(\Psi)-f(\Psi))}\ell^{\mu\rho}\partial_\rho\Psi\ell^{\nu\sigma}\partial_\sigma\Psi, \label{pmdt3}
\end{equation}
where one can easily see that they have the desired property of being invariant under a Weyl transformation $\ell_{\mu\nu}\rightarrow\Omega^2(x)\ell_{\mu\nu}$.
For simplicity, we assume that the original scalar-tensor theory is actually just Einstein's General Relativity and in the following we also assume that the contribution of the baryons to the expansion can be neglected. The equations of motion of this model, Eqs. (\ref{kinematicalc}) and (\ref{EOM_mimetic1}), are then
\begin{eqnarray}
G_{\mu\nu}=G b(\Psi) \partial_\mu\Psi\partial_\nu\Psi,\qquad 2 \nabla_\rho\left(Gb\partial^\rho\Psi\right)=G\frac{1}{b(\Psi)}\frac{db(\Psi)}{d\Psi},\qquad b(\Psi)g^{\mu\nu}\partial_\mu\Psi\partial_\nu\Psi=1,\label{eomDR}
\end{eqnarray}
where $G$ denotes the trace of the Einstein tensor $G_{\mu\nu}$ as $G=g^{\mu\nu}G_{\mu\nu}$.
We will look for cosmological solutions by setting the metric $\ell_{\mu\nu}$ to be equal to a flat FLRW metric and $\Psi$ to be a function of time only. This implies that the non-zero components of the metric $g_{\mu\nu}$ (there is isotropy so the y and z component are equal to the x component) are
\begin{equation}
g_{tt}=b\dot\Psi^2,\quad g^{tt}=1/(b\dot\Psi^2),\quad g_{xx}=a^2 A,\quad g^{xx}=1/(a^2 A).\label{metriccomponents}
\end{equation}
The tt and xx components of the previous generalized Einstein equations are equal to each other and equal to
\begin{eqnarray}
&&4bA\left[A(3H^2+2\dot H)\dot\Psi-2AH\ddot\Psi+\dot\Psi^2(A_{,\Psi\Psi}-4A_{,\Psi w}\ddot\Psi+4A_{,ww}\ddot\Psi^2)+\dot\Psi^2(3HA_{,\Psi}-2 A_{,w}(\dddot\Psi+3H\ddot{\Psi}))\right]
\nonumber\\&&
-2A\dot\Psi^2\left(2AH+\dot\Psi(A_{,\Psi}-2A_{,w}\ddot\Psi)\right)b'=b\dot\Psi^3(A_{,\Psi}-2A_{,w}\ddot\Psi)^2,\label{eomgEE}
\end{eqnarray}
where prime means derivative with respect to $\Psi$ and $A_{,\Psi}$, $A_{,w}$ and so on denote derivatives of $A$ with respect to $\Psi$ and $w$ respectively. Eq. (\ref{eomgEE}) contains higher-derivatives for $\Psi$ and they disappear if $A$ does not depend on $w$. Note that from the kinematical constraint in (\ref{eomDR}) we do not get any equation of motion if it is written in terms of the metric $\ell_{\mu\nu}$. One can also show that by combining Eq. (\ref{eomgEE}) with its derivative one can find the equation of motion for $\Psi$.

Now let us consider a particular mimetic disformal transformation of the type (\ref{pmdt1}), (\ref{pmdt2}) and (\ref{pmdt3}). In this case we have Weyl invariance and we can choose to fix the gauge as
\begin{equation}
A(\Psi,w)=(b(\Psi)-f(\Psi))w=1.\label{mygauge}
\end{equation}
In this new gauge, the independent generalized Einstein equation is
\begin{equation}
(3H^2+2\dot H)\dot\Psi-2H\ddot\Psi=H\dot\Psi^2\frac{b'}{b}.
\end{equation}
By doing a change of variables as
\begin{equation}
\frac{d\Phi}{d\Psi}=\sqrt{|b|},\label{changeofvariables}
\end{equation}
one can simplify the equation to
\begin{equation}
(3H^2+2\dot H)\dot\Phi-2H\ddot\Phi=0.\label{eomnew}
\end{equation}
The equation of motion (\ref{eomnew}) can be written as
$
\frac{d(\ln \dot\phi)}{dt}=\frac{3}{2}\frac{d(\ln a)}{dt}+\frac{d(\ln H)}{dt}
$.
So one can integrate it once to find $\dot\Phi(t)=\mathrm{const}\,a(t)^\frac{3}{2}H(t)$,
where $\mathrm{const}$ denotes a constant of integration. And integrating once more to find
\begin{equation}
\Phi(t)=\mathrm{const}_1a(t)^\frac{3}{2}+\mathrm{const}_2,\label{phiofa}
\end{equation}
where $\mathrm{const}_i$ with $i=1,2,3,4$ denote constants of integration.
Using Eq. (\ref{metriccomponents}), the change of variables (\ref{changeofvariables}) and our gauge choice $A=1$, the $g_{\mu\nu}$ metric components can be written as
\begin{equation}
g_{tt}(t)=\mathrm{Sign}(b)\dot\Phi(t)^2,\qquad g_{xx}(t)=a^2(t).
\end{equation}
By doing a change of time variable as
\begin{equation}
\sqrt{-g_{tt}(t)}dt=d\tilde t
\end{equation}
one can find that
\begin{equation}
a(t(\tilde t))=(\mathrm{const}_3 \tilde t +\mathrm{const}_4)^\frac{2}{3},
\end{equation}
which is a matter dominated universe for the $g_{\mu\nu}$ metric. This is always the case for any $\ell_{\mu\nu}$, $b(\Psi)$ or $f(\Psi)$. This result is consistent with the original findings of \cite{Chamseddine:2013kea} that one only gets a matter universe as a solution for the metric $g_{\mu\nu}$ having started from mimetic General Relativity by doing a conformal transformation\footnote{If $f=0$ then the disformal transformation Eq. (\ref{pmdt2}) becomes simply a conformal transformation. Furthermore, the gauge condition Eq. (\ref{mygauge}) implies that $g_{\mu\nu}=\ell_{\mu\nu}$ and that $\dot\Phi$ is a constant. This singles out the matter-dominated universe solution from the set of all possible solutions of Eq. (\ref{phiofa}). This case is nothing more than the result of \cite{Chamseddine:2013kea}.}. This result is expected in light of the findings of \cite{Deruelle:2014zza} that showed that the mimetic dark matter equations of motion, Eqs. (\ref{eomDR}), when written in terms of $g_{\mu\nu}$ are the same for any mimetic disformal transformation. However now in our case the physical metric is $\ell_{\mu\nu}$ so from Eq.  (\ref{phiofa}) one can see that in this model we can have any expansion history desired. For a given scale factor solution $\ell_{xx}=a^2(t)$ the scalar field $\Phi$ adjusts according to Eq.  (\ref{phiofa}) in order for this to be a solution of the equation of motion Eq. (\ref{eomnew}). The solution for the original field $\Psi$ can be found once we specify the function $b(\Psi)$ by using Eq.  (\ref{changeofvariables}). Finally the function $f(\Psi)$ is found by using the gauge condition, Eq. (\ref{mygauge}), which in the background can be written as $1=-\dot\Psi^2(b-f)=-\dot\Phi^2(\mathrm{Sign}(b)-f/|b|)$. In other words, by choosing a specific function $f$ one can obtain the desired scale factor solution.
For instance, de Sitter spacetime is a solution of (\ref{eomnew}) for $\Phi(t)\propto e^{3H/2t}$. A matter universe solution results from taking $\dot\Phi=\mathrm{constant}$. The expansion history of a universe filled with a barotropic perfect fluid with a constant equations of state $\omega$ and a cosmological constant $\Lambda$ is $a=a_\star\sinh^{\frac{2}{3(1+\omega)}}\left(\mathcal{C}t\right)$,
where $\mathcal{C}=\sqrt{3\Lambda/4}(1+\omega)$. The solution for $\Phi$ is $\Phi(t)=C_2+C_1 \sinh^\frac{1}{1+\omega}(\mathcal{C}t)$, where $C_1$ and $C_2$ are integration constants.

\section{Conclusions \label{sec:CON}}

In this paper, we showed that a very general scalar-tensor theory of the type (\ref{STgeneralaction}) is generically invariant under a disformal tansformation of the kind (\ref{distrans}), irrespective of whether the scalar field in the action is the same or different than the scalar field involved in the transformation. We also showed that there is a special subset of those disformal transformations for which the previous result is not valid. We call those special disformal transformations as mimetic disformal transformations because they give origin to a new scalar-tensor theory of gravity that is a generalization of the ``mimetic" dark matter proposal \cite{Chamseddine:2013kea}. These mimetic disformal transformations are given by Eq. (\ref{mimetic B}). They basically are a subset of (\ref{distrans}) where the two free functions $A(\Psi,X)$ and $B(\Psi,X)$ are related as (\ref{mimetic B}). These results generalize the findings of \cite{Deruelle:2014zza} that were obtained for Einstein's General Relativity. We also showed that the reason why a simple change of variables as in a mimetic disformal transformation leads to a new physical theory is because we are doing a non-invertible change of variables. If the change of variables is invertible then the physical theory does not change as expected. The derived non-invertibility or mimetic condition is the same for any general scalar-tensor theory, as it is the property of the disformal tansformation of the kind of Eq. (\ref{distrans}).
We have shown that the mimetic equations of motion of the new scalar-tensor theory can be derived from an action containing an extra scalar field playing the role of a Lagrange multiplier that imposes the kinematical constraint (\ref{normalizationcondition}) to be satisfied throughout the dynamics. Again this generalizes some results in \cite{Golovnev:2013jxa,Barvinsky:2013mea} to a general scalar-tensor theory context.

As an application of some of our findings, we have presented a simple toy model of the mimetic Horndeski theory where a canonically normalized scalar field with no potential (in the original theory) can be used to mimic the background expansion history of a universe filled with a barotropic perfect fluid with a constant equations of state. Actually, we showed that in this simple scalar-tensor model one can have almost any (the restriction is that the effective pressure cannot change sign) desired background expansion history by suitably choosing the ``potential" function $b(\Psi)$ in the action (\ref{generalaction}). We have generalized the previous simple model to include a cubic Galileon interaction and as an example we showed that this model can easily mimic the background expansion history of a universe filled with dark matter and a positive cosmological constant.

We also presented an example where instead of minimally coupling baryons with $g_{\mu\nu}$ we coupled them minimally with the new metric $\ell_{\mu\nu}$. In this case, for the original theory, we took simply the disformally transformed Einstein-Hilbert action. We again found that, for a cosmological background, the metric $\ell_{\mu\nu}$ can have any expansion history desired by suitably choosing the free functions $b(\Psi)$ and $f(\Psi)$ in the mimetic disformal transformation (\ref{pmdt2}).

Finally, we also showed that the mimetic theory, when written in terms of the metric $g_{\mu\nu}$, does not contain any more derivatives than the scalar-tensor theory that originated it. This may not be the case for the mimetic theory when written in terms of the new metric $\ell_{\mu\nu}$.

\section{Acknowledgments}

We would like to thank Bin Hu and Luigi Pilo for discussions. We are also grateful to Shahab Shahidi and Alexander Vikman for comments on the first version of the manuscript and for drawing our attention to Refs. \cite{Lim:2010yk,Mirzagholi:2014ifa,Haghani:2015iva}. FA is supported by the National Taiwan University (NTU) under Project No. 103R4000 and by the NTU Leung Center for Cosmology and Particle Astrophysics (LeCosPA) under Project No. FI121. PK acknowledges financial support from a Cariparo foundation grant.

\appendix
\section{\label{GDT}Disformal transformation with a new scalar field}

In Section \ref{sec:DIS} of the main text we have considered the theory that results from performing a disformal transformation on a very general scalar-tensor theory where the scalar field in the action was the same as the scalar field involved in the transformation. In this Appendix, we consider the case when the scalar field in the transformation is not the same as the scalar field present in the action of the theory.

The action of the model is Eq. (\ref{STgeneralaction}). The disformal transformation that we are considering is
\begin{equation}
g_{\mu\nu}=A(\Phi,Y)\,\ell_{\mu\nu}+B(\Phi,Y)\,\partial_\mu\Phi\partial_\nu\Phi,\quad\hbox{where}\quad Y\equiv\ell^{\rho\sigma}\partial_\rho\Phi\partial_\sigma\Phi,\label{disformed metric}
\end{equation}
and, as before, the arbitrary disformal functions, $A$ and $B$, depends on both the scalar field $\Phi$ and its kinetic term $Y$.
Using Eq. (\ref{disformed metric}) one can find the variation of $g_{\mu\nu}$ as
\begin{eqnarray}
\delta g_{\mu\nu}&=&A\,\delta\ell_{\mu\nu}-\left(\ell_{\mu\nu}{\partial A\over\partial {Y}}+\partial_\mu\Phi\partial_\nu\Phi{\partial B\over \partial {Y}}\right)\left[(\ell^{\alpha\rho}\partial_\alpha\Phi)\,
(\ell^{\beta\sigma}\partial_\beta\Phi)\,\delta\ell_{\rho\sigma}-2\ell^{\rho\sigma}(\partial_\rho\Phi)\,(\partial_\sigma\delta\Phi)\right]\nonumber\\
&&+\left(\ell_{\mu\nu}{\partial A\over\partial  \Phi}+\partial_\mu\Phi\partial_\nu\Phi{\partial B\over \partial \Phi}\right)\delta\Phi +B\left[(\partial_\mu\Phi)(\partial_\nu\delta\Phi)+(\partial_\nu\Phi)(\partial_\mu\delta\Phi)\right]. \label{g_variation2}
\end{eqnarray}

The modified Einstein equations of motion, $\delta S/\delta\ell_{\mu\nu}=0$, are
\begin{eqnarray}
 A(E^{\mu\nu}+T^{\mu\nu})&=&\left(\alpha_1{\partial A\over\partial {Y}}+\alpha_3{\partial B\over\partial {Y}}\right)(\ell^{\mu\rho}\partial_\rho\Phi)\,
(\ell^{\nu\sigma}\partial_\sigma\Phi). \label{new_E_eom_2}
\end{eqnarray}
The equation of motion for the scalar field $\Phi$, $\delta S/\delta\Phi=0$, is
\begin{equation}
{1\over\sqrt{-g}}\partial_\rho\left\{\sqrt{-g}\,\partial_\sigma\Phi\left[B(E^{\rho\sigma}+ T^{\rho\sigma})+\left(\alpha_1{\partial A\over\partial {Y}}+\alpha_3{\partial B\over\partial {Y}}\right)\ell^{\rho\sigma}\right]
\right\}=
{1\over2}\left(\alpha_1{\partial A\over\partial\Phi}+\alpha_3{\partial B\over\partial\Phi}\right),\label{new_psi_eom_2}
\end{equation}
where we have defined the new quantities $\alpha_1$ and $\alpha_3$ as
\begin{equation}
 \alpha_1 \equiv (E^{\rho\sigma}+ T^{\rho\sigma})\ell_{\rho\sigma},\quad\quad \alpha_3\equiv(E^{\rho\sigma}+ T^{\rho\sigma})\partial_\rho\Phi\,\partial_\sigma\Phi,
\end{equation}
and the equation for the scalar field $\Psi$, $\delta S/\delta\Psi=0$, is $\Omega_\Psi = 0$.
For the matter field we have the same equation of motion as before, i.e. $\delta S/\delta\phi_m=0$ or $\Omega_m=0$.
It is important to note that the modified Einstein equations of motion, Eq. (\ref{new_E_eom_2}), have the same structure as in subsection \ref{subsec:DISM} of the main text. Following the same procedure as before one can find different solutions for the resulting system depending on its determinant being zero or not. We will now consider these two cases separately and show that they lead to different physical theories exactly as in the case studied in the main text.

\subsubsection{The generic case}

If the determinant of the system of linear equations that results from contracting Eq. (\ref{new_E_eom_2}) with $\ell_{\mu\nu}$ and $\partial_\mu\Phi\partial_\nu\Phi$ is non-zero then the only solution is $\alpha_1=\alpha_3=0$. Hence, Eqs. (\ref{new_E_eom_2}) and (\ref{new_psi_eom_2}) imply
\begin{eqnarray}
A(E^{\mu\nu}+T^{\mu\nu})&=&0, \label{recover_E_eqn_1}\\
\partial_\rho\left[\sqrt{-g}\,\partial_\sigma\Phi B(E^{\rho\sigma}+ T^{\rho\sigma}) \right] &=& 0. \label{recover_psi_eom_2}
\end{eqnarray}
Eq. (\ref{recover_psi_eom_2}) is empty after considering the modified Einstein equation (\ref{recover_E_eqn_1}). Therefore the full equations of motion of this theory are
\begin{eqnarray}
E^{\mu\nu}+T^{\mu\nu}=0, \quad \Omega_\Psi=0, \quad \Omega_m=0.\label{appendaux}
\end{eqnarray}
In terms of the original metric $g_{\mu\nu}$, these equations are exactly the same as the equations of motion for the theory (\ref{STgeneralaction}) if we take the variation with respect to the original fields $g_{\mu\nu}$, $\Psi$ and $\phi_m$ instead of taking the variation with respect to the new fields $\ell_{\mu\nu}$, $\Phi$, $\Psi$ and $\phi_m$ as we did to arrive at Eqs. (\ref{appendaux}). This shows that a very general scalar-tensor theory of the type (\ref{STgeneralaction}) is invariant under a generic disformal transformation of the type (\ref{disformed metric}) even if the scalar field defining the transformation in not the same as the scalar field in the action.

\subsubsection{Mimetic gravity}

If the determinant of the system in zero, as we showed in the main text, it implies that the transformation functions $A$ and $B$ should be related as
\begin{equation}
B(\Phi,{Y})= -{A(\Phi, {Y})\over {Y}}+b(\Phi),\label{mimetic B2}
\end{equation}
where $b(\Phi)$ again arises as an integration constant in $Y$ which we assume to be non-zero.
Inserting this expression in Eqs. (\ref{new_E_eom_2}) and (\ref{new_psi_eom_2}) one obtains
\begin{equation}
E_{\mu\nu}+ T_{\mu\nu}=(E+T)\,b\,\partial_\mu\Phi\,\partial_\nu\Phi,\quad\quad
\nabla_\rho\left[(E+T)b\,\partial^\rho\Phi\right] = \frac{1}{2}(E+T) {1\over b}{db\over d\Phi}.
\end{equation}

Following \cite{Deruelle:2014zza}, one can redefine the scalar field $\Phi$ in terms of a new scalar field $\Theta$ as $d\Theta/d\Phi=\sqrt{|b|}$ in order to eliminate the function $b(\Phi)$ from the equations of motion to finally obtain
\begin{equation}
E_{\mu\nu}+ T_{\mu\nu}=\epsilon(E+T)\partial_\mu\Theta\,\partial_\nu\Theta,\quad\quad
\nabla_\rho\left[(E+T)\partial^\rho\Theta\right]=0,\label{mimetic eom2}
\end{equation}
where $\epsilon=\mathrm{Sign}(b)=\pm1$ depending on the sign of the norm of $\partial_\mu\Phi$, i.e. $g^{\mu\nu}\partial_\mu\,\Theta\partial_\nu\Theta=\epsilon$.
The full set of equations of motion includes in addition to the previous two equations also the equations of motion for $\Psi$ and the matter field $\phi_m$. They are
\begin{equation}
\Omega_\Psi=0, \qquad \Omega_m=0.
\end{equation}


\end{document}